\documentclass[letterpaper, 10 pt, conference]{ieeeconf}
\IEEEoverridecommandlockouts
\usepackage{caption}
\renewcommand{\thetable}{\arabic{table}}
\usepackage{booktabs}
\usepackage{array}
\usepackage{diagbox}    
\usepackage{cite}
\usepackage{xurl}
\usepackage{float}
\usepackage{amsmath,amssymb,amsfonts}
\usepackage{algorithmic}
\usepackage{graphicx}
\usepackage{textcomp}
\usepackage{xcolor}
\usepackage{soul}
\usepackage{theorem}
\def\BibTeX{{\rm B\kern-.05em{\sc i\kern-.025em b}\kern-.08em
    T\kern-.1667em\lower.7ex\hbox{E}\kern-.125emX}}
\usepackage{algorithm}
\usepackage{algorithmic}
\newtheorem{remark}{Remark}

\usepackage{hyperref}
\hypersetup{pdfborder={0 0 0},}

\begin{document}



\title{\LARGE \bf ALADIN-$\beta$: A Distributed Optimization Algorithm for Solving MPCC Problems
}

\author{Yifei Wang, Shuting Wu, Genke Yang, Jian Chu, Apostolos I. Rikos and Xu Du$^*$ 
\thanks{$^*$ Corresponding author.}
\thanks{Yifei Wang, Genke Yang and Jian Chu are with the Ningbo Artificial Intelligence Institute and the Department of Automation, Shanghai Jiao Tong University, Ningbo 315000, China, and also with the School of Automation and Intelligent Sensing, Shanghai Jiao Tong University, Shanghai 200240, China. E-mails: \texttt{\{yifeiw4ng, gkyang, chujian\}@sjtu.edu.cn}.}
\thanks{Shuting Wu is with School of Mathematics and Statistics, North China University of Water Resources and Electric Power, Zhengzhou, China, and is also with the HNAS Institute of Mathematics, Henan Academy of Science, Zhengzhou, China. E-mail: \texttt{wushuting0126@163.com}.}
        \thanks{Apostolos I. Rikos and Xu Du are with the Artificial Intelligence Thrust of the Information Hub, The Hong Kong University of Science and Technology (Guangzhou), Guangzhou, China. 
    Apostolos I. Rikos is also affiliated with the Department of Computer Science and Engineering, The Hong Kong University of Science and Technology, Clear Water Bay, Hong Kong, China. E-mails: \texttt{\{apostolosr, michaelxudu\}@hkust-gz.edu.cn}.}
 \thanks{The work of A.I.R. and X.D. are supported by the Guangzhou-HKUST(GZ) Joint Funding Scheme (Grant No. 2025A03J3960).}}


\maketitle

\begin{abstract}
Mathematical Programs with Complementarity Constraints (MPCC) are critical in various real-world applications but notoriously challenging due to non-smoothness and degeneracy from complementarity constraints. 
The $\ell_1$-Exact Penalty-Barrier enhanced \texttt{IPOPT} improves performance and robustness by introducing additional inequality constraints and decision variables. However, this comes at the cost of increased computational complexity due to the higher dimensionality and additional constraints introduced by the centralized formulation. 
To mitigate this, we propose a distributed structure-splitting reformulation that decomposes these inequality constraints and auxiliary variables into independent sub-problems. 
Furthermore, we introduce Augmented Lagrangian Alternating Direction Inexact Newton (ALADIN)-$\beta$, a novel approach that integrates the $\ell_1$-Exact Penalty-Barrier method with ALADIN to efficiently solve the distributed reformulation. 
Numerical experiments demonstrate that even without a globalization strategy, the proposed distributed approach achieves the fast convergence while maintaining high precision. 

\end{abstract}


\section{Introduction}

Mathematical Programs with Complementarity Constraints (MPCC) \cite{leyffer2003mathematical} are a class of optimization problems incorporating complementarity constraints. They have become a valuable modeling framework in control theory, particularly for optimization-based control problems involving non-smooth dynamical systems \cite{Nurkanovic2023f} and hybrid systems \cite{van1998complementarity}. Key challenges in optimal control, such as trajectory optimization for contact-aware robotic manipulators \cite{katayarna2022quasistatic}, logic-based control in chemical processes \cite{kazi2025optimal}, and mixed Boolean optimization for smart grids \cite{du2023optimal}, can be systematically reformulated as MPCC through a unified framework \cite{nurkanovic2024solving}, thereby leveraging complementarity structures to encode non-smooth physical and logical constraints. To address degeneracies in such problems, several solution methods have been proposed \cite{kim2020mpec}. Notably, a method was introduced that bypasses direct solution of the original problem by utilizing sequential convex programming to solve penalty sub-problems derived from linear complementarity quadratic programs, with penalty parameters adaptively adjusted through homotopy \cite{Hall2021}, \cite{Hall2024}. Building on these foundations, a subsequent study extends the original method to tackle a general mixed-integer nonlinear problem arising from antenna selection optimization in a sequential quadratic programming (SQP) fashion \cite{zhu2022alternating}. Additionally, \cite{thierry2020ℓ1} applied an $\ell_1$-exact penalty strategy to handle complementarity constraints, coupled with the penalty parameter update strategy to address degenerate problems with \texttt{IPOPT}. This integration demonstrates improved speed and robustness on standard MPCC benchmarks by introducing additional inequality constraints and decision variables. However, as a centralized approach, this $\ell_1$-strategy inherently struggles with increased computational complexity due to higher dimensionality and additional constraints. As problem sizes grow, memory constraints and processing bottlenecks further limit its scalability. This motivates the exploration of distributed optimization (DO) for efficient MPCC solving.

DO algorithms have become essential for tackling large-scale engineering problems \cite{Nedic2018,Engelmann2019,Du2019,boyd2011distributed} by mitigating the \emph{curse of dimensionality} through decomposition into parallelizable sub-problems, improving scalability and reducing communication overhead. They typically follow two main approaches: primal and dual decomposition \cite{ling2015dlm}. 
Dual decomposition methods, particularly Alternating Direction Method of Multipliers (ADMM) \cite{boyd2011distributed} and Augmented Lagrangian Alternating Direction Inexact Newton (ALADIN) \cite{Houska2016,Du2025,du2025convergence,du2023consensus}, offer rigorous theoretical convergence guarantees and exhibit practical effectiveness in DO.
Theoretically, ALADIN synthesizes the distributed coordination efficiency of ADMM with the high precision of SQP, guaranteeing global convergence for convex problems \cite{Houska2021} and local convergence for non-convex problems \cite{Houska2016}. Recent advancements like ALADIN-Prox \cite{du2023bi}, integrate proximal regularization with ALADIN, extending this to global convergence in non-convex settings with preserved complexity. In practice, ALADIN exhibits exceptional performance in large-scale engineering applications, primarily categorized into three types: a) \emph{spatial-splitting-based} optimization problems (e.g., optimal power flow) \cite{Engelmann2019,Du2019,engelmann2020decomposition}, b) \emph{time-splitting-based} optimization problems (e.g., model predictive control (MPC) and moving horizon estimation (MHE)) \cite{Kouzoupis2016,9815081,stomberg2022decentralized,wu2025time}, and c)
the distributed reconstruction of non-summation centralized optimization \cite{Shi2022}, also known as \emph{structure-splitting-based} optimization.
The remarkable numerical performance of ALADIN relies on the fulfillment of regularity conditions such as the Linear Independence Constraint Qualification (LICQ) and Second-Order Sufficiency Conditions (SOSC), yet these conditions are inherently violated in degenerate problems like MPCC. Specifically, MPCC often violate standard constraint qualifications (CQ) at every feasible point, due to the non-convexity and non-smoothness introduced by the complementarity constraints. To the best of our knowledge, existing research on ALADIN has not addressed MPCC. This work proposes an enhanced ALADIN framework tailored for MPCCs, aiming to bridge this methodological gap.

Building on the $\ell_1$-penalty method \cite{thierry2020ℓ1} and the structure-splitting-based optimization of ALADIN \cite{Shi2022}, we propose a novel DO algorithm, ALADIN‑$\beta$, for solving MPCC problems. 
First, following the approach in \cite{thierry2020ℓ1}, the $\ell_1$-exact penalty method is applied to handle the non-smooth and non-convex complementarity constraints, and then a universally applicable structure-splitting paradigm is proposed for the resulting general nonlinear programming (NLP) problems, see Section \ref{sec: reformulation}. Next, we propose ALADIN-$\beta$ for solving the distributed problems derived from the structure-splitting process, see Section \ref{sec: Algorithm section}. Finally, representative numerical examples are compared with existing state-of-the-art solvers to validate the effectiveness and efficiency of ALADIN‑$\beta$, see Section \ref{sec: numerical}.

\section{Preliminaries}\label{sec: Preliminary}

This section begins with a review of NLP with orthogonality constraints. We then revisit the foundation of the $\ell_1$-exact penalty-barrier-based interior-point method, an efficient centralized approach for solving MPCC. Finally, we provide an overview of the ALADIN algorithm, a powerful algorithm designed for large-scale distributed optimization.

\subsection{Problem Formulation}\label{sec: Formulation}
An NLP with orthogonality constraints can be formulated as follows,\vspace{-3mm}
\begin{subequations}\label{eq:general NLP}
\begin{align}
\min_{x\in\mathbb{R}^n} & \quad f(x) \\
\text{s.t.}\hspace{1mm} & \quad g(x) = 0, \label{eq: MPCC}\\
&\quad x \leq 0\label{eq: inequality}.
\end{align}
\end{subequations}
Here, $f: \mathbb{R}^n \rightarrow \mathbb{R}$ and $g: \mathbb{R}^n \rightarrow \mathbb{R}$ are all assumed to be twice continuously differentiable. Moreover, equation~\eqref{eq: MPCC} defines the orthogonality constraints of the form $g(x) = G(x)^\top H(x)$, where $G:\mathbb{R}^n \rightarrow \mathbb{R}^{n_g}$, $H:\mathbb{R}^n \rightarrow \mathbb{R}^{n_g}$ are twice continuously differentiable. For simplicity, this work focuses on linear inequality constraints. However, leveraging the ALADIN framework \cite{Houska2016}, future extensions will address more general inequality constraints.


\begin{remark}
MPCC are 
considered as a class of degenerate NLPs, where redundant constraints violate CQs\footnote{Common CQs include LICQ and the weaker Mangasarian-Fromovitz Constraint Qualification (MFCQ) \cite{Nocedal2006}. While LICQ (sufficient but not necessary) ensures the existence of bounded and unique Lagrange multipliers by requiring linearly independent active constraint gradients, the weaker MFCQ guarantees only the boundedness of the multipliers. Despite being less stringent, MFCQ plays a crucial role in the convergence of most NLP solvers due to its wider applicability in degenerate problems.}. Typically, the complementarity constraints in MPCC, formulated as $0\leq G(x)\perp H(x)\geq0$, can be equivalently expressed via smooth constraints: $G(x)^\top H(x)=0$, $G(x)\geq 0$ and $H(x)\geq 0$. Active set methods resolve small-scale degeneracies effectively, yet their combinatorial complexity escalates with dimensionality. Conversely, interior-point methods like \texttt{IPOPT} \cite{wachter2006implementation} regularize constraints globally under localized degeneracy assumptions but struggle with numerical instability due to inconsistent constraint linearizations. 

\end{remark}


\subsection{The $\ell_1$-Exact Penalty-Barrier-Based 
Reformulation}\label{sec: ifac l1-exact penalty-barrier reformulation}
To tackle the aforementioned challenges in solving problem~\eqref{eq:general NLP}, we review an improved interior-point method in this section. This method systematically mitigates degeneracy-induced issues while preserving numerical robustness across problem scales.

By incorporating the inequality constraint \eqref{eq: inequality} into the objective function using barrier terms, we formulate a sequence of barrier problems whose solutions asymptotically converge to those of \eqref{eq:general NLP} as the barrier parameter $\mu$ approaches zero\footnote{Under the MFCQ, the sequence of local minimizers for the barrier-augmented function asymptotically converges to the optimal solution $x^{\star}$ of \eqref{eq:general NLP} as $\mu\rightarrow 0$.}.
 To further penalize the equality constraint \eqref{eq: MPCC}, 
 we yield the combined non-smooth $\ell_1$-penalty term\footnote{To relax the restrictive LICQ, we reformulate problem~\eqref{eq:general NLP} into an exact $\ell_1$-penalty-barrier framework, where the equivalence to the original problem is preserved if $\rho \geq \rho^{\star} := \| (\lambda^{\star},z^{\star})\|_{\infty}$ with $\lambda^{\star},z^{\star}$ denoting optimal dual variables \cite{Nocedal2006}.} with a parameter $\rho$,\vspace{-1mm}
\begin{equation}\label{eq:l1-penalty-barrier nonsmooth}\small
\begin{aligned}
\underset{x\in\mathbb{R}^n}{\text{minimize}} \quad f(x) - \mu \sum_{i = 1}^{n} \ln (-x_{i}) + \rho \|g(x)\|_1,
\end{aligned}
\end{equation}
where $\mu\in \mathbb{R}_{>0}$ is the barrier parameter and $\rho\in \mathbb{R}_{>0}$ is the penalty parameter. 
However, the non-smoothness induced by the $\ell_1$-penalty term complicates the minimization process, necessitating specialized non-smooth optimization strategies.

To address this, \cite{thierry2020ℓ1} introduces auxiliary variables $p$ and $n$ to augment the equality constraints, resulting in a smooth reformulation of problem 
\eqref{eq:l1-penalty-barrier nonsmooth}:\vspace{-1mm}
\begin{equation}\label{eq:l1-penalty-barrier smooth}\small
\begin{aligned}
 \min_{x\in\mathbb{R}^{n};p,n\in\mathbb{R}^{m}} & \quad f(x)- \mu \sum_{i = 1}^{n} \ln (-x_{i})+\rho (p+n)^\top e \\
  \text{s.t.}\hspace{5.5mm}& \quad g(x)-p+n=0,\quad p\geq0,\quad n\geq0,
\end{aligned}
\end{equation}
where $e = [1,1,\cdots,1]^\top\in \mathbb R^m$ denotes a vector of appropriate dimension filled with ones. Crucially, the reformulated equality constraints maintain full-rank properties, ensuring the LICQ holds even at rank-deficient points
of the barrier problem. This enables robust computation of steps for degenerate problems. 
Replacing the proposed $\ell_1$-exact penalty-barrier phase in the \texttt{IPOPT} solver has been shown to significantly improve success rates for degenerate problems and accurately solve a practical MPCC problem in a single optimization stage \cite{thierry2020ℓ1}. 
This approach avoids numerical challenges associated with non-smooth terms but comes at the cost of increased problem dimensions and the need for an additional penalty parameter update strategy.


\subsection{Fundamentals of ALADIN for QP}\label{sec: classical ALADIN}
In this section, we review how ALADIN, a DO algorithm, handles large-scale QPs with inequality constraints \cite{Shi2022}.  Before presenting ALADIN, we define the local update notation: $(\cdot)^+$ denotes the value after the update, while $(\cdot)^-$ represents the value before the update.

Considering a strictly convex quadratic programming problem 
\begin{equation}\label{eq:general NLP2}\small
\begin{aligned}
\min_{x\in\mathbb{R}^n} & \quad \tilde{f}(x) \\
\text{s.t.}\hspace{1mm} &\quad \underline{x}\leq x \leq \overline{x},
\end{aligned}
\end{equation}
with $\tilde{f}(x) = \frac{1}{2}x^\top Q x$, where $Q\succ 0$, $\underline{x}$ and $\overline{x}$ denote the lower and upper bound of $x$, respectively.
By introducing an auxiliary variable $z$, the inequality constraints of problem 
\eqref{eq:general NLP2} can be incorporated into the objective function in a reformulated manner,
\begin{equation}\label{eq: ALADIN ref}
\begin{aligned}
\min\limits_{x, z} \quad \tilde{f}(x) + \tilde{g}(z)\quad \text{s.t.} \quad x = z \mid \lambda,
\end{aligned}
\end{equation}
where the function \vspace{-0.5mm}
\begin{equation}
\tilde{g}(z)\overset{\text{def}}{=}\left\{
    \begin{split}
        0\quad & \text{if} \quad\underline{z}\leq z \leq \overline{z}\\
         \infty \quad & \text{otherwise}
    \end{split}\right\}
\end{equation}
is defined as the indicator function, with $\underline{z} = \underline{x}$ and $\overline{z} = \overline{x}$, and $\lambda$ denotes the dual variable with respect to the coupling constraint. The customized ALADIN \cite{Shi2022} for solving problem~\eqref{eq:general NLP2} initializes primal-dual variables $(x,z,\lambda)$ and selects positive-definite matrices $\Sigma\succ 0$ and $K\succ 0$, proceeding as follows,\vspace{0mm}
\begin{equation}\label{framework: aladin acc ver}
\begin{aligned}
\begin{cases}
v^+ = \arg\min\limits_{v} \tilde{f}(v) + \lambda^\top  v + \frac{1}{2} \|v - x\|^2_\Sigma; \\
w^+ = \arg\min\limits_{w} \tilde{g}(w) - \lambda^\top w+ \frac{1}{2} \|w - z\|^2_K; \\
\quad\left( x^+, z^+, \lambda^+ \right) \\
\hspace{-0.5mm}= \hspace{-1mm}\begin{cases}
\min\limits_{x^+, z^+} 
\hspace{-3mm}&\frac{1}{2} \left\| \begin{bmatrix} x^+ - v^+ \\ z^+ - w^+ \end{bmatrix} \right\|_{H,K}^2 
\hspace{-2mm}+ \hspace{-1mm}\begin{bmatrix} \hspace{-0.5mm}\nabla \tilde{f}(v^+) \\ \hspace{1mm} \partial \tilde{g}(w^+) \hspace{-0.5mm}\end{bmatrix}^\top \hspace{-1mm}
\begin{bmatrix} x^+ \\ z^+ \end{bmatrix}
\\
\hspace{2mm}\text{s.t.} \hspace{-2mm}&  x^+ = z^+ \mid \lambda^+.
\end{cases}
\end{cases}
\end{aligned}
\end{equation}
Here, $\nabla \tilde{f}$ and $\partial \tilde{g}$ denote the gradient and subgradient of $\tilde{f}$ and $\tilde{g}$, respectively, while the matrices $H\succ 0$ and $K\succ 0$ are local Hessian approximations of $\tilde{f}$ and $\tilde{g}$. Note that, ALADIN alternates between parallel sub-problems solving at sub-nodes to update \(v^+\) and \(w^+\), and coordination at the central node to determine \((x^+, z^+, \lambda^+)\), iterating until convergence. 
This study demonstrates the significant potential of ALADIN in solving large-scale problems, focusing on solving convex QPs \cite[Section IV-B]{Shi2022}. Its extension to non-convex optimization problems will be presented in subsequent sections.


\section{Structure-Splitting-Based Reformulation}\label{sec: reformulation}
Building on Sections \ref{sec: ifac l1-exact penalty-barrier reformulation} and \ref{sec: classical ALADIN}, we introduce a structure-splitting reformulation of problem~\eqref{eq:l1-penalty-barrier smooth} by incorporating auxiliary variables. The optimization variables are grouped into independent blocks, resulting in a generalized representation of the structure-splitting formulation as follows.
 \vspace{-2mm} 
\begin{subequations}\label{eq:distributed reformulation1}
\begin{align}
 \min_{\alpha, \beta, \gamma} &\quad  \phi(\alpha,\gamma) + \varphi(\beta) + \psi(\gamma)\label{eq:objective function distributed}\\
 \text{s.t.}\hspace{1mm}& \quad g(x)-q+m=0\label{local_eq_constraints}, \\
 &\quad x = \beta,\quad m = n, \quad p= q.
\end{align}
\end{subequations} 
The composite variables are structured as $\alpha = [x^\top, q^\top, m^\top]^\top$ and $\gamma = [p^\top, n^\top]^\top $, where $q$ and $ m $ act as local replicas of the auxiliary variables $p$ and $n$, respectively, to enforce consensus among distributed sub-problems. Furthermore, the auxiliary variable $\beta$ represents the local variables associated with the log-barrier term for the inequality constraints in \eqref{eq:l1-penalty-barrier smooth}. To enhance numerical stability, a relaxation parameter $r$ is integrated into the formulation, strategically designed to mitigate abrupt gradient-driven oscillations near constraint boundaries. Furthermore, the objective function \eqref{eq:objective function distributed} is decomposed into three distinct components, expressed as follows,
\vspace{-1mm}\begin{subequations}
\label{eq: subproblems}
    \begin{align}
       & \phi(\alpha,\gamma) \overset{\text{def}}{=}f(x) + \frac{1}{2}\left\| q-p \right\|^2_P+\frac{1}{2}\left\| m-n \right\|^2_M, \label{eq: first}\\
        & \varphi(\beta) \overset{\text{def}}{=}    -\sum_{i = 1}^{n} \mu_i\ln (r-\beta_i) ,\label{eq: second}\\
         &\psi(\gamma) \overset{\text{def}}{=} \hspace{-0.5mm}  \rho (p+n)\hspace{-0.5mm}^\top \hspace{-0.5mm}e \hspace{-0.5mm}- \hspace{-0.5mm} \sum_{i = 1}^{n_g}\hspace{-0.5mm}\mu_i \hspace{-0.5mm}\left(\ln (r\hspace{-0.5mm}+\hspace{-0.5mm}p_i)\hspace{-0.5mm} + \hspace{-0.5mm}\ln (r\hspace{-0.5mm}+\hspace{-0.5mm}n_i)\right)\label{eq: third},
    \end{align}
\end{subequations}
with $r>0$, $\mu_i>0$, $P\succ0$, $M\succ 0$. 
Sub-problem \eqref{eq: first} is augmented by two proximal terms in $q$ and $m$ to regularize the update step sizes near optimality. Inspired by \cite{Shi2022}, the inequality constraints in \eqref{eq:l1-penalty-barrier smooth} are reformulated as \emph{relaxed log-barrier} objectives, eliminating the need for explicit feasibility detection. Refer to \cite{Shi2022} for details.

By introducing the coupling matrices\vspace{-1mm}
\begin{equation}
\begin{split}
&A_1 \overset{\text{def}}{=} \begin{bmatrix}
        I_{|x|} & \mathbf 0_{|x|\times n_g} &\mathbf 0_{|x|\times n_g}\\
        \mathbf 0_{n_g\times|x|} &I_{n_g} & \mathbf 0_{n_g\times n_g}\\
        \mathbf 0_{n_g\times|x| } &\mathbf 0_{n_g\times n_g} & I_{n_g}
    \end{bmatrix},\\
     &   A_2 \overset{\text{def}}{=} \begin{bmatrix}
        -I_{|x|} \\
        \mathbf 0_{n_g\times|x|} \\
        \mathbf 0_{n_g\times|x|}
    \end{bmatrix},\quad
      A_3 \overset{\text{def}}{=} \begin{bmatrix}
         \mathbf 0_{|x|\times n_g} &\mathbf 0_{|x|\times n_g}\\
        -I_{n_g} & \mathbf 0_{n_g\times n_g}\\
    \mathbf 0_{n_g\times n_g} & -I_{n_g}
    \end{bmatrix},
    \end{split}
\end{equation}
Problem~\eqref{eq:distributed reformulation1} can then be represented as,
\begin{equation}\label{eq:distributed reformulation2}
\begin{aligned}
 \min_{\alpha, \beta, \gamma} &\quad  \phi(\alpha,\gamma) + \varphi(\beta) + \psi(\gamma)\\
\hspace{-4mm} \text{s.t.}\hspace{1mm}& \quad G(\alpha)=0\hspace{23.5mm} | \kappa, \\
 &\quad A_1 \alpha + A_2 \beta +A_3\gamma = 0\hspace{3.9mm} | \lambda,
\end{aligned}
\end{equation} 
where $G(\alpha) = g(x)-q+m$, $\kappa$ represents the dual variable of the first constraint and $\lambda$ denotes the dual variable of the coupling constraint.


\vspace{-1.5mm}
\begin{remark}
Drawing inspiration from \eqref{eq: ALADIN ref}, we employ an analogous technique to reformulate the inequality-constrained problem 
\eqref{eq:l1-penalty-barrier smooth} into a distributed reformulation \eqref{eq:distributed reformulation2} by incorporating inequality constraints into the objective function. This reformulation strategically eliminates the computational overhead associated with infeasibility detection in constrained optimization. Note that, although the formulation of problem~\eqref{eq:distributed reformulation1} introduces new variables and a modified objective function, its summation structure enables distributed solving. 


\end{remark}

\section{{ALADIN}-$\beta$}\label{sec: Algorithm section}
In order to solve the reformulated distributed $\ell_1$-penalty-barrier problem 
\eqref{eq:distributed reformulation2}, we develop an efficient solution approach leveraging the ALADIN algorithm and detail its implementation strategy.

\subsection{Algorithm Development}\label{sec: ALADIN algorithm}
We propose a novel algorithm, ALADIN-$\beta$, to solve \eqref{eq:distributed reformulation2}. The algorithm begins with an initial guess for all primal and dual variables $(\alpha,\beta,\gamma ,\lambda)$ in \eqref{eq:distributed reformulation2}. It then iterates through the following steps until the termination criteria are met. The main steps of ALADIN-$\beta$ are outlined below.



\noindent
1) Choose proper initial values for the barrier and penalty parameter $\mu, \rho \geq 0$ and positive semi-definite scaling matrices $\Sigma_i\in\mathbb{S}_{+}^{n_x}$ for all $i\in\{1,2,3\}$.

\noindent
2)
Solve the decoupled optimization problems in parallel,
\begin{itemize}
    \item For Sub-problem 1:\vspace{-1mm}\begin{equation}\label{eq:sub-problem 1}
\begin{aligned}\vspace{-0.5mm}
 \hat\alpha\hspace{-0.5mm}=\hspace{-0.5mm}\arg\min_{\alpha}\hspace{-2mm} & \quad  \phi(\alpha,\gamma^-)\hspace{-1mm} + \hspace{-1mm}\lambda^\top A_1 \alpha\hspace{-0.5mm} + \hspace{-0.5mm}\frac{1}{2} \|\alpha - \alpha^{-}\|^2_{\Sigma_1}\\
  \text{s.t.} \hspace{-2mm}& \quad G(\alpha) = 0 \mid \kappa.
\end{aligned}
\end{equation} 
    \item For Sub-problem 2: \vspace{-1mm}\begin{equation}\label{eq:sub-problem 2}
    \begin{aligned}
 \hat\beta\hspace{-0.5mm}=\hspace{-0.5mm}\arg\min_{\beta}\hspace{-2.5mm} \quad  \varphi(\beta)\hspace{-0.5mm}+\hspace{-0.5mm} \lambda^\top A_2 \beta \hspace{-0.5mm} + \hspace{-0.5mm}\frac{1}{2} \|\beta - \beta^{-}\|^2_{\Sigma_2}.
\end{aligned}
\end{equation}    \vspace{-1mm}
    \item For Sub-problem 3: 
    \vspace{-3mm}\begin{equation}\label{eq:sub-problem 3}
 \hat\gamma\hspace{-0.5mm}=\hspace{-0.5mm}\arg\min_{\gamma}\hspace{-2.5mm} \quad  \psi(\gamma) \hspace{-0.5mm} + \hspace{-0.5mm}\lambda^\top A_3 \gamma\hspace{-0.5mm} + \hspace{-0.5mm}\frac{1}{2} \|\gamma- \gamma^{-}\|^2_{\Sigma_3}.
\end{equation}
\end{itemize}
\noindent
3) Evaluate the sensitivity information for each sub-problem, obtaining the gradient $g_i$ and Hessian $H_i$ for all $i\in\{1,2,3\}$ based on their respective local optimal solutions, expressed as follows, 
\vspace{-1mm}\begin{subequations}
\label{eq:sub-problem 2 gradient}
\begin{align}
\nabla _{\alpha}\phi(\hat\alpha) & = 
\Sigma_1(\alpha^{-}-\hat \alpha)-A_1^\top \lambda - \nabla _{\alpha}G(\hat \alpha)^\top \kappa,\\
\nabla_{\beta} \varphi(\hat \beta) & = \Sigma_2(\beta^{-}-\hat \beta)-A_2^\top\lambda,\\
\nabla_{\gamma} \psi(\hat\gamma) & = 
\Sigma_3(\gamma^--\hat \gamma) -A_3^\top \lambda.
\end{align}
\end{subequations}
Then choose symmetric Hessian approximations 
\vspace{-1mm}\begin{subequations}\label{eq:sub-problem 2 hessian}
\begin{align}
H_1 &\approx \nabla^2_{\alpha}\{\phi(\hat \alpha,\gamma^-) + \kappa^\top G(\hat \alpha)\}\succ 0,\\\vspace{-5mm}
H_2 &\approx \nabla^2_{\beta}\{\varphi(\hat\beta)\} \hspace{-0.5mm}=\hspace{-0.5mm} \text{diag}\left(\frac{\mu_i }{(r-\hat \beta_i)^2}\right) \hspace{-0.7mm}\succ\hspace{-0.7mm}0,\\
H_3 &\approx \nabla^2_{\gamma}\{\psi(\hat\gamma)\} \hspace{-0.5mm}=\hspace{-0.5mm}\text{diag}\left(\hspace{-0.7mm}\frac{\mu_i}{(r-\hat p_i)^2}\hspace{-0.7mm}+\hspace{-0.7mm}\frac{\mu_i}{(r-\hat n_i)^2}\hspace{-0.7mm}\right)\hspace{-0.7mm}\succ\hspace{-0.7mm}0.
\end{align}
\end{subequations}
\noindent
4) Aggregate the sensitivity information from all sub-problems to construct and solve the consensus QP problem subject to equality constraints,\footnote{For QP problems with equality constraints, we can derive closed-form solutions for updating the primal and dual variables using the first-order KKT conditions. This allows the primal updates to be distributed to the sub-nodes for internal updating, thereby further enhancing computational efficiency for large-scale problems.} 
\vspace{1mm}\begin{equation}\label{eq:ALADIN consensus QP}
\begin{aligned}
 \min_{\alpha^+,\beta^+,\gamma^+} &\hspace{2mm}\frac{1}{2} \left\| \begin{bmatrix} \alpha^+ - \hat\alpha \\ \beta^+ -\hat\beta\\\gamma^+ -\hat\gamma\end{bmatrix} \right\|_{H}^2 
\hspace{-1mm}+ \begin{bmatrix} \nabla_{\alpha}\phi(\hat\alpha) \\ \hspace{1mm} \nabla_{\beta}\varphi(\hat\beta) \\ \hspace{1mm} \nabla_{\gamma}\psi(\hat\gamma) \end{bmatrix}^\top \begin{bmatrix}
    \alpha^+- \hat\alpha \\ \beta^+ - \hat\beta\\ \gamma^+ - \hat\gamma
\end{bmatrix}
\\
 \text{s.t.}\hspace{4mm} & \quad C \ (\alpha^+ - \hat\alpha)  = 0,\\
 & \quad A_1\alpha^+ +A_2\beta^+ +A_3\gamma^+  = 0 \mid \lambda^{\text{QP}}.
\end{aligned}
\end{equation}
Here the matrix $H$ is a block diagonal matrix composed of $ H_i, \forall i\in\{1,2,3\}$ from \eqref{eq:sub-problem 2 hessian}. 
Moreover, $C= \nabla_{\alpha} G(\alpha)^\top=\left(\frac{\partial g(x)}{\partial x}, -I_{n_g}, I_{n_g}\right)$ denotes the Jacobian matrix of the local equality constraints.

\noindent
5)
Update the primal variables $\alpha^{+}\rightarrow\alpha^{-}$, $\beta^{+}\rightarrow\beta^{-}$,  $\gamma^{+}\rightarrow\gamma^{-}$, as well as $\lambda = \lambda +\theta(\lambda^{\text{QP}}-\lambda)$ with $\theta = 1$ for a full-step 
update\footnote{For simplicity, only the full-step update is presented here. If the original problem is highly nonlinear or the initial point is far from the optimal solution, techniques such as line search or other globalization strategies can be employed to improve convergence performance, as detailed in \cite{Houska2016, du2023bi}.}. Verify whether the stopping conditions of the algorithm are satisfied. If not, update the barrier parameter $\mu_j, \rho$ and go to Step $2$); otherwise, terminate the iteration.
\vspace{-2mm}\begin{remark}\label{remark3}
The proposed algorithm {ALADIN}-$\beta$ systematically decomposes the original problem into independent sub-problems, enabling parallel computation. 
Sub-problem \eqref{eq:sub-problem 1} utilizes the values of \( p \) and \( n \) from the previous iteration to solve the primary objective and equality constraints. As detailed in Section 
\ref{sec: ifac l1-exact penalty-barrier reformulation}, its constraints are reformulated to eliminate dependence on the LICQ.
This property allows ALADIN-$\beta$ to avoid various numerical issues that arise when solving systems with degeneracies. 
The subsequent two sub-problems 
(see \eqref{eq:sub-problem 2}, \eqref{eq:sub-problem 3}) are introduced to iteratively refine the results from Sub-problem 1. Specifically, Sub-problem \eqref{eq:sub-problem 2} enforces inequality constraints through log-barrier penalty terms, while Sub-problem \eqref{eq:sub-problem 3} adaptively adjusts $p$ and $n$ to tighten equality constraint satisfaction. These auxiliary sub-problems, designed with inequality-constrained log-barrier formulations, exhibit simplified structures ensuring analytical closed-form solutions and guaranteed positive definiteness of their Hessian matrices. 
Moreover, incorporating proximal terms for $q$ and $m$ in the objective function of Sub-problem \eqref{eq: first} not only improves its convexity but also enhances the stability of algorithm convergence across iterations.
\end{remark}
\vspace{-4mm}\begin{remark}\label{remark4}

    Based on the reformulation in Section~\ref{sec: ifac l1-exact penalty-barrier reformulation}, ALADIN‑$\beta$ extends standard ALADIN by relaxing locally LICQ and SOSC requirements \cite{Houska2016, Du2025, du2025convergence}, improving robustness to degeneracies and enabling application to MPCCs. It retains ALADIN’s distributed structure by decomposing the problem into parallelizable sub-problems and solving a simplified high-dimensional QP. In strongly coupled yet structurally simple problems, ALADIN‑$\beta$ outperforms the $\ell_1$-exact penalty-barrier phase of interior-point methods in distributed settings. Our smoothing-based MPCC reformulation \cite{thierry2020ℓ1} broadens ALADIN’s applicability while preserving its local convergence properties. A full convergence analysis will be provided in an extended version of this work.
\end{remark}


\section{Numerical Experiment}\label{sec: numerical}

This section presents simulation results demonstrating the advantages of ALADIN-$\beta$ for MPCC problems and compares it with two variants of the $\ell_1$-penalized \texttt{IPOPT} solver \cite{thierry2020ℓ1}, referred to as \texttt{IPOPT1} and \texttt{IPOPT2}\footnote{ In ALADIN $\beta$, the parameters $\rho$
and $\mu$ are updated at each iteration. In contrast, \texttt{IPOPT2} \cite{thierry2020ℓ1} updates these parameters only after solving the current barrier problem to optimality for fixed values of $\rho$ and $\mu$. This fundamental discrepancy in parameter update strategies renders a direct performance comparison between the two methods inherently unfair.
To establish a more equitable basis for comparison, we introduce \texttt{IPOPT1}, which follows the algorithmic framework proposed in \cite{thierry2020ℓ1} but incorporates parameter updates for $\rho$ and $\mu$ at every iteration, thereby aligning its update scheme with that of ALADIN--$\beta$.}, as well as a baseline \texttt{IPOPT}-vanilla that directly solves the original MPCC problem \eqref{eq:general NLP} without reformulation.
Notably, we emphasize the QP step, as ALADIN tolerates inexact subproblem solutions without compromising convergence, leading to reduced computational effort supported by theoretical analysis \cite{du2025convergence} and numerical evidence \cite{wu2025time}.
In \texttt{IPOPT1}, the penalty parameter $\mu$ is updated after a single QP step, while in \texttt{IPOPT2}, it is updated only after fully solving the current barrier problem. The implementation of ALADIN-$\beta$ is based on \texttt{Casadi-v3.6.7} with \texttt{IPOPT} and \texttt{MATLAB 2024b}.
Moreover, to eliminate the influence of heuristic tuning, all the implemented algorithms update the parameters at a fixed rate, where the barrier parameter $\mu =10$ (see \eqref{eq: subproblems}) decreases by a factor of $0.2$, and the penalty parameter $\rho = 10$ increases by a factor of $4$ at each update.

To clearly demonstrate the numerical performance of ALADIN-$\beta$, we adopt a canonical MPCC example (see page $21$ of \url{https://www.syscop.de/files/2023ss/nonsmooth_school/kirches_MPECs_2.pdf}), presented as follows:
\begin{equation}\label{eq:numerical example}
\begin{aligned}
\min_{\hat{x},\tilde{x}\in\mathbb{R}^n} & \quad \frac{1}{2}\|\hat{x}-e\|_2^2+\frac{1}{2}\|\tilde{x}-e\|_2^2 \\
\text{s.t.}\hspace{2mm} &\quad \hat{x}^\top\tilde{x} = 0,\\
&\quad x\succeq 0.
\end{aligned}
\end{equation}
Here, $\hat{x}$ and $\tilde{x}$ are n-dimensional column vectors, with $x=[\hat{x}^\top,\tilde{x}^\top]^\top$. Despite the convexity of the objective function and inequality constraints, the equality constraints introduce non-convexity and non-smoothness, violating CQs and posing challenges for standard numerical algorithms.

\begin{figure}[h]
\centering
\includegraphics[width=0.34\textwidth,height=0.24\textheight]{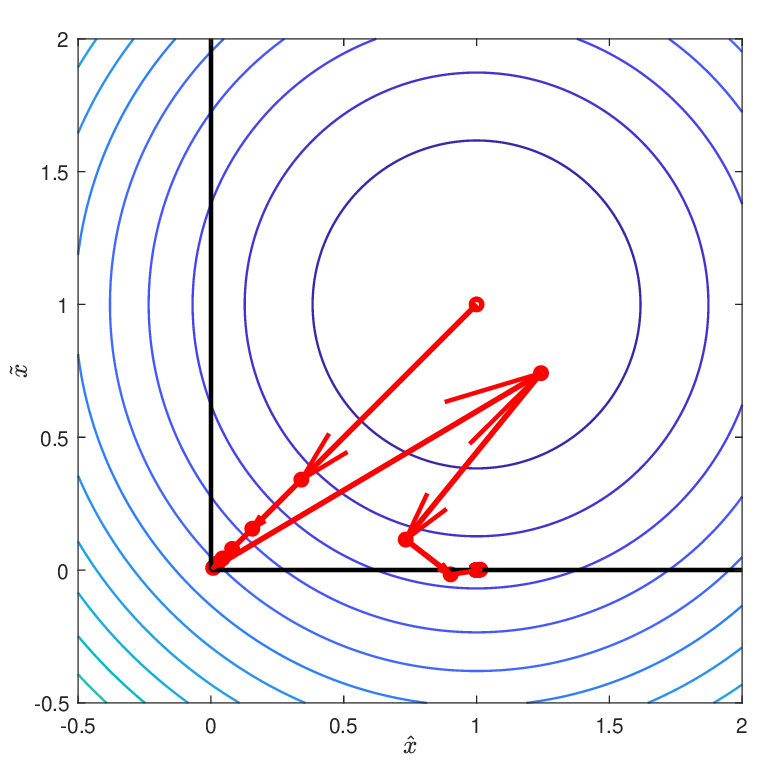}
\caption{Iteration process of ALADIN-$\beta$ on the two-dimensional instance of problem~\eqref{eq:numerical example}.}
\label{fig: contour plot}
\end{figure}

Figure \ref{fig: contour plot} illustrates the proposed ALADIN-$\beta$ algorithm solving problem~\eqref{eq:numerical example}, where $\hat{x},\tilde{x}\in \mathbb R_+$. The problem features two local minima at $(0,1)$ and $(1,0)$, along with a relatively weak local maximum at the origin.

Here, we fix the total number of ALADIN-$\beta$ iterations to $100$ and set the initial point to $(1,1)$. The trajectory first moves toward the origin to satisfy the complementarity constraints and, once these are nearly met, pivots toward objective minimization, ultimately converging to the local optimum $(1,0)$ of \eqref{eq:numerical example}.

\begin{figure}[h]
\centering
\includegraphics[width=0.48\textwidth,height=0.43\textheight]{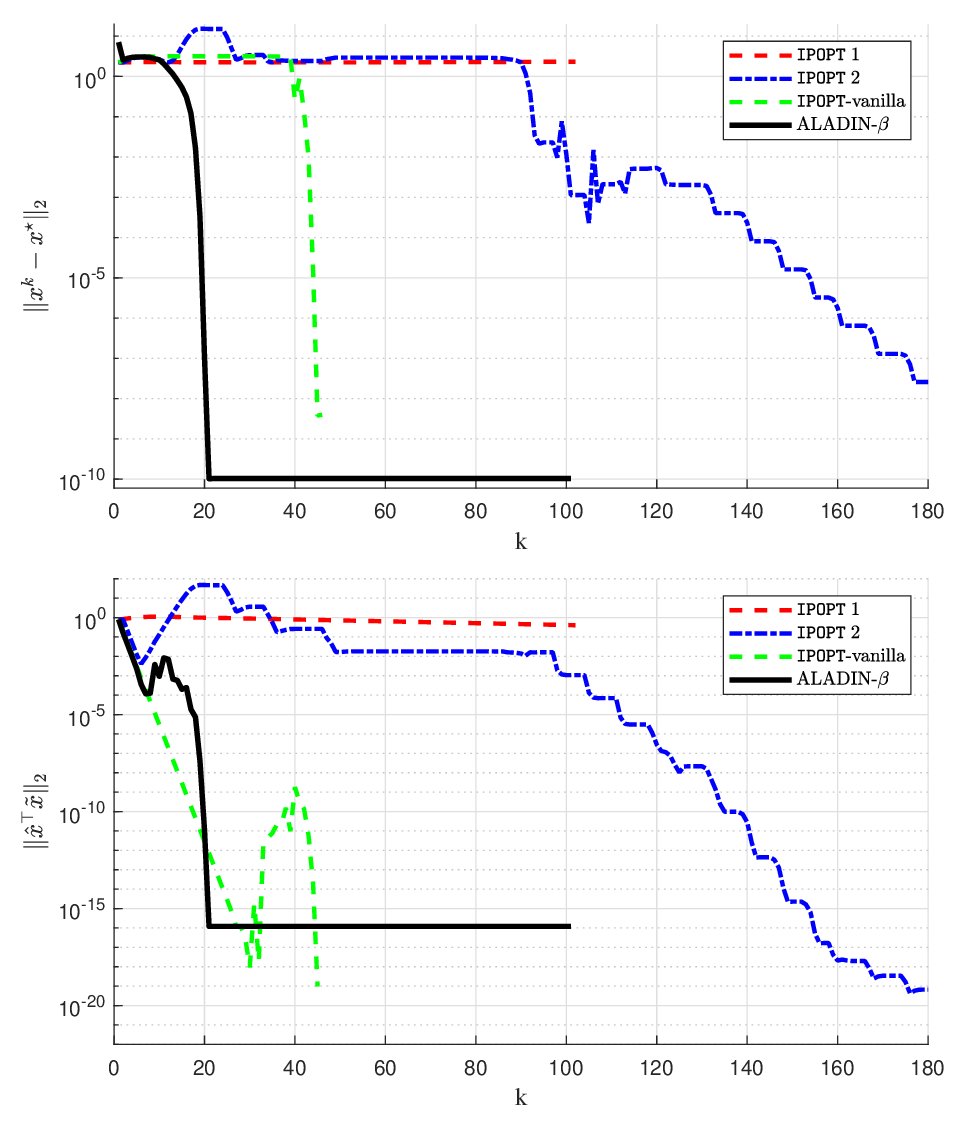}
\caption{Convergence comparison among ALADIN-$\beta$ and two types of \texttt{IPOPT}.}
\label{fig: convergence plot}
\end{figure}

Figure \ref{fig: convergence plot} presents the convergence behavior of \texttt{IPOPT1}, \texttt{IPOPT2}, \texttt{IPOPT}-vanilla, and ALADIN-$\beta$ for solving the MPCC problem \eqref{eq:numerical example} with $\hat{x},\tilde{x}\in \mathbb R^{10}$. The upper subplot shows the convergence of the primal variable $x$. \texttt{IPOPT2} exhibits slow convergence in the first 90 iterations, followed by linear convergence to a precision of $10^{-8}$ after approximately 180 iterations. In contrast, ALADIN-vanilla and ALADIN-$\beta$ both exhibit quadratic convergence, attaining $10^{-9}$ accuracy in $42$ iterations and $10^{-10}$ accuracy within the fewest iterations, respectively.
The lower subplot tracks the complementarity constraint residuals of \eqref{eq:numerical example}. ALADIN-$\beta$ satisfies these constraints to $10^{-16}$ precision within approximately $20$ iterations, equiring the fewest iterations among all methods.
Notably, \texttt{IPOPT}-based methods incorporate adaptive globalization strategies (such as the filter method) to maintain convergence beyond $10^{-16}$ precision, achieving higher accuracy at a consistent convergence rate compared to ALADIN-$\beta$.
Despite both \texttt{IPOPT1} and ALADIN-$\beta$ updating parameters after each QP step, the convergence curves of \texttt{IPOPT1} in Figure \ref{fig: convergence plot} show slow convergence.


In DO, the way local information is exchanged and utilized typically limits the ability of the algorithm to achieve higher convergence precision. However, as illustrated in Figure \ref{fig: convergence plot}, the proposed ALADIN-$\beta$, by appropriately utilizing the first- and second-order information as equation~\eqref{eq:sub-problem 2 gradient} and \eqref{eq:sub-problem 2 hessian}, converges to an acceptable accuracy of $10^{-10}$ in the fewest iterations, outperforming all the \texttt{IPOPT}-based methods. Notably, as demonstrated in the presentation of ALADIN (see \url{https://www.uiam.sk/~oravec/apvv_sk_cn/slides/aladin.pdf}, page $61$), the ALADIN algorithm significantly reduces CPU time per iteration for MPC problems compared to conventional numerical methods \cite{Kouzoupis2016}. Since ALADIN-$\beta$ inherits the structure of the standard ALADIN (see Remark \ref{remark3}), it also holds the potential to reduce the computational time required to solve MPCC problems.

\section{Conclusion}
This work presents ALADIN-$\beta$, a novel distributed algorithm that integrates exact-penalty methods to efficiently and robustly solve MPCC problems in parallel. By systematically avoiding failures caused by violated CQs in ill-conditioned problems, ALADIN-$\beta$ relaxes the LICQ requirements of classical ALADIN, expanding its applicability. Numerical experiments on a benchmark MPCC demonstrates its superior convergence speed and accuracy compared to state-of-the-art NLP solver variants. Future work will focus on formalizing convergence proofs, incorporating globalization strategies, and developing adaptive parameter tuning to enhance robustness for complex MPCCs.




    \bibliographystyle{IEEEtran}
\bibliography{paper}
		
\end{document}